# On the optimality of the standard genetic code: the role of stop codons

Sergey Naumenko[1*], Andrew Podlazov[1], Mikhail Burtsev[1,2], George Malinetsky[1]

[1]*Department of Non-linear Dynamics, Keldysh Institute for Applied Mathematics of RAS, Moscow 125047, Russia.*

[2]*Department of Systemogenesis, Anokhin Institute of Normal Physiology of RAMS, Moscow 125009, Russia.*

Corresponding author: Sergey Naumenko

Address for Correspondence: Russian Academy of Sciences, Keldysh Institute for Applied Mathematics, Department of Non-linear Dynamics, Miusskaya Place 4, RU-125047, Moscow, Russia

E-mail:sergey_clark@yahoo.co.uk

Tel.: +7 (495) 2507802

Fax: +7 (495) 9720737

**Abstract**

The genetic code markup is the assignment of stop codons. The standard genetic code markup ensures the maximum possible stability of genetic information with respect to two fault classes: frameshift and nonsense mutations. There are only 528 (about 1,3% of total number) optimal markups in the set of markups having 3 stop codons. Among the sets of markups with 1,2…8 stop codons, the standard case having 3 stop codons has maximum absolute number of optimal markups.

## Introduction

The standard genetic code is shared by all living organisms with a few insignificant exceptions. Formally, the genetic code is a mapping of an alphabet consisting of 64 codons onto a set consisting of 20 letters (amino-acids) and one punctuation mark.

Amino acids are coded in genome by triplets of nucleotides. The position of a nucleotide in a triplet is significant. Therefore, there are $4^3$=64 different codons. There are 61 triplets among the codons, which encode amino acids and 3 stop codons which terminate the protein synthesis process. The choice of 3 stop codons constitutes the *genetic code markup*.

The standard genetic code is one of many possibilities (Trainor, 2001). Even before the complete decryption of the genetic code there was a question: *why* amino-acids are coded just this way (Woese, 1965). For a long time the natural genetic code was thought to be a «frozen accident» (Crick, 1968). But many statistical studies support the theory that this genetic code has evolved towards minimizing errors of transcription and translation (Goodarzi et al., 2004). In particular, it has been proved that natural genetic code minimizes the effect of point mutations or mistranslations: either the erroneous codon is a synonym of the original amino acid, or it encodes an amino acid with similar chemical properties (Freeland, Hurst, 1998). Apart from this fact the natural genetic code possesses a set of symmetries and a semantic structure (Gusev,Shulze-Makuch, 2004).

**The main goal of our work is to find out *why* the genetic code uses TAA, TAG and TGA codons as punctuation marks.**

A choice of stop codons affects error protection of encoded information in case of frameshift and point mutations.

**Frameshift mutations**.

A codon is entirely defined by the starting position of triplet reading or the reading frame. Therefore there are 3 different ways to read the same nucleic sequence depending on reading frame shift (Fig. 1). Below we shall call the gene reading with left shift by 1 nucleotide as ‘shift 1’, and the reading with right shift by 1 nucleotide as ‘shift 2’ (Fig. 2).

If a pair of consecutive sense codons gives stop codon in process of reading with a shift, we call it a *terminating pair*.

*Optimization task 1* consists in minimization of the influence of frameshift mutations due to maximizing the number of terminating pairs of sense codons.

**Point mutations**.

Point mutation in a sense codon may result in appearance of sense codon or stop codon (Fig. 3.), i.e. the markup affects the probability of nonsense mutations. Point mutations leading to the transformation of a sense codon into a stop one are named *nonsense* mutations. We name codons for which nonsense mutation is possible as *vulnerable* codons. The total number of nonsense mutations (over the entire code) is equal or greater than a number of vulnerable codons because a vulnerable codon may be subjected to several different nonsense mutations.

*Optimization task 2-a* consists in minimization of the number of vulnerable codons.
*Optimization task 2-b* consists in minimization of the number of nonsense mutations.

We assume that the genetic code has a protection mechanism on the level of its markup, i.e. on the level of stop codons choice, contrary to the biochemical level.

The first group of questions we address is related to optimization task 1:

1) How does the choice of stop codons affect blocking of frameshift mutations?
2) What values can possess the number of terminating pairs of sense codons?
3) How is the set of genetic code markups with 3 stop codons distributed according to the possible values of the number of terminating pairs of sense codons?
4) How optimal is the canonical markup from the point of view of optimization task 1?

The second group of questions is related to optimization task 2:

1) How does the choice of stop codons affect blocking of point mutations?
2) What values can possess the number of vulnerable codons and the number of nonsense mutations?
3) How is the set of genetic code markups with 3 stop codons distributed according to the possible values of the number of vulnerable codons and the number of nonsense mutations?
4) How optimal is the canonical markup from the point of view of optimization task 2?

Is the canonical markup optimal for task 2-a or for task 2-b?

All these questions are related to markups with 3 stop codons. Finally, it is interesting, how many optimal markups exist in sets with various numbers of stop codons?

**Methods**

Crick et al. considered a set of codes involving nonoverlapping triplets of nucleotides. Each triplet codes one amino acid. All codes have no stop codons. Crick et al. showed that to avoid frameshift mutations, we must limit the number of different kinds of amino acids that the code can handle. They proved that the upper bound equals 20 and showed that a code for 20 amino acids exists (Crick et al, 1957). It is well known that the experimentally found number equals 20 and this research is an example of the power of simple genetic code models.

In a number of statistical studies (see review in Goodarzi et al., 2004) the canonical genetic code is compared with randomly generated codes in order to assess relative efficiency of the natural code with the various types of fitness functions.

Our approach is closer to the former (Crick et al.,1957) one rather than the latter.

At first we consider *only markups with 3 stop codons*. There are $C_{64}^{3} = \frac{64!}{61!3!} = \frac{64 \cdot 63 \cdot 62}{1 \cdot 2 \cdot 3} = 41664$ such markups. For every markup from this set we calculate the following values:

1) the number of terminating pairs of codons in the case of shift 1;
2) the number of terminating pairs of codons in the case of shift 2;
3) the number of vulnerable codons;
4) the number of missense mutations.

We work with characteristics 1-2 and 3-4 separately. For example, there are markups with the same characteristics 1. We group markups with the same values of the number of terminating pairs of codons in the case of shift 1 and calculate their number. After the detailed study of the set of markups containing 3 stop codons, we apply the same algorithm to the sets with 1,2, … 8 stop codons.

**Results**

The choice of stop codons affects the number of terminating pairs of sense codons. The bigger is the number of terminating pairs in a genetic code markup, the better it blocks the frameshift mutations.

Let us consider different genetic code markups with 3 stop codons, and for each code calculate the maximum possible number of terminating pairs of codons for both types of shift (left or right). There are 61×61=3721 pairs of sense codons. They form sequences of 6 characters belonging to the {A;G;C;T} alphabet. Neither the first nor the second part of such sequence (the first part consists of characters 1-2-3, the second one consists of characters 4-5-6) is a stop codon. The pair of sense codons is terminating one if it contains 3 sequential nucleotides forming a stop codon in the positions 2–3–4 or 3–4–5 (Fig.2.).

The maximum possible number of terminating pairs of codons for each type of shift equals 192=3×4×4×4 (3 variants of stop-codon × 4×4×4 variants of another nucleotides). This maximum is reached in shift 1 and shift 2 simultaneously. Indeed, necessary and sufficient condition for the markup's optimality in opmization task 1 is the absence of stop codons coinciding first and last nucleotides (different codons, for example TA**G** and **G**CT or T**AG** and **AG**C, or single codon, for example **CCC** or **G**T**G)**.

The calculation shows, that the canonical markup has the maximum possible number of terminating pairs for both frame shifts. There are 2432 markups (this is 5,8% of total number of markups (41664)) with 192 terminating pairs for both frame shifts. The distribution of markups vs the number of terminating pairs of codons for the shift 1 (for the shift 2 the distribution is the same) is shown in Fig.4.

A choice of stop codons affects the number of nonsense mutations and vulnerable codons. There are 27 possible point mutations leading to one of the 3 stop codons (3 stop-codons × 3 triplet positions × 3 nucleotide types). Every point mutation, which transforms a stop codon into another stop codon, decreases the number of nonsense mutations. The maximum number of mutations transforming a stop codon into another one equals 6.

All possible markups fall into 10 classes according to the number of vulnerable codons and to the number of nonsense mutations. It is possible to combine these 10 classes into 4 groups (a-d) according to the number of nonsense mutations for each class (see Table 1).

In the group (a) one stop codon can be transformed to any of two others, but those cannot be transformed to each other by point mutation. The standard genetic code belongs to this group (standard stop codons: TAA, TAG, TGA). Here the total number of point mutations transforming stop codon to stop codon equals 4. Consequently, for the sense part of code there are 27 – 4 = 23 nonsense mutations. In this case, 5 sense codons can be transformed to one of two stop codons and 23-5×2 = 13 sense codons can be transformed to only one stop codon. Hence, there are 5+13 = 18 vulnerable codons in total. This is the minimum possible value of vulnerable codons among all 10 classes. It is interesting that the standard genetic code markup has the minimum number of vulnerable codons rather than the minimum number of nonsense mutations.

In the group (b) for every stop codon there are 2 point mutations transforming it to another stop codon. There are 6 such mutations. In this group only one of sense codons can be transformed to any of 3 stop codons and 6 sense codons can be transformed to one of 3 stop codons. Hence, there are 1+6+6+6 = 19 vulnerable codons and 1×3+6+6+6 = 21 possible nonsense mutations.

In the group (c) there are only two possible point mutations transforming a stop codon to a stop codon. Only one stop codon mutates to only one of others two and vice versa. Hence, there are 25 nonsense mutations and 20 or 21 or 23 vulnerable codons in this group.

In the group (d) none of stop-codons can be transformed to another one. All of 27 nonsense mutations belong to the sense part of the code. A number of vulnerable codons may be equal 21, 22, 23, 25, 27.

We found that there are 2432 markups with the minimum possible number of sense codons which can be transformed to stop codon by point mutation, and only 528 of them (about 1,3% of the total number of markups) have the maximum number of terminating pairs. The canonical markup belongs to this small set.

If one assumes that the genetic code markup evolves, then not only stop codons assignments should be varied but the number of stop codons itself should be changed too. We considered the sets of markups with 1,2,… 8 stop codons and calculated numbers of optimal markups (with highest numbers of terminating pairs and lowest numbers of vulnerable codons). Figure 5 summarizes our calculations. The case with 3 stop codons (528 optimal markups) is the local and the global maximum. (See also Table 2).

**Thus we proved that the choice of stop codons TAA, TAG and TGA in the standard genetic code ensures the protection of information encoded with respect to the frameshift and nonsense mutations.**

Detailed results published in (Naumenko, Podlazov, 2005).

**Discussion**

Having these results we can hypothesize that the following factors affect the evolution of the genetic code markup. (i) The shift of the reading frame is a non-local error which leads to completely different sequence of codons. Therefore the corresponding sequence of amino-acids would also be changed completely. This unexpected new protein would have totally different properties and function. It completely distorts the “meaning” of the gene. In this case not only the cell resources are spent on synthesizing of nonfunctional protein, but the resulting amino acid sequence may be harmful. The best way to handle this problem is to stop the mutant gene expression as soon as possible. Therefore, the genetic code markups with higher number of terminating pairs protect a cell better against possible damage caused by frameshift mutation. (ii) An accidental substitution of a nucleotide in a sense codon leads to transformation of a sense codon into another sense one or into a stop one. In the first case mutation can be silent, i.e. the resulting protein will not be changed at all, or, even if new amino acid is different, the protein may preserve its functionality due to locality of modification in protein structure. Therefore, the mutant protein would probably have similar properties as the original one and would perform its functions correctly. In the case of nonsense mutation, i.e. transformation from sense codon to stop one, the initial gene sequence will be truncated. As a result, the protein will lose its functional properties with high probability. Hence, minimizing the number of vulnerable codons in the genetic code markup makes a cell more robust to point mutations.

Our results indicate that among all genetic code markups with 3 stop codons the standard markup has the maximum possible probability of terminating gene reading process in the case of frameshift mutation and minimal number of sense codons which can be transformed to a stop one by point mutation. Thus, in its class the standard markup assures the best protection against possible damage in the cases of frameshift and point mutations. These findings support the general hypothesis that the genetic code is not a frozen accident but on the contrary, it is a result of evolution.

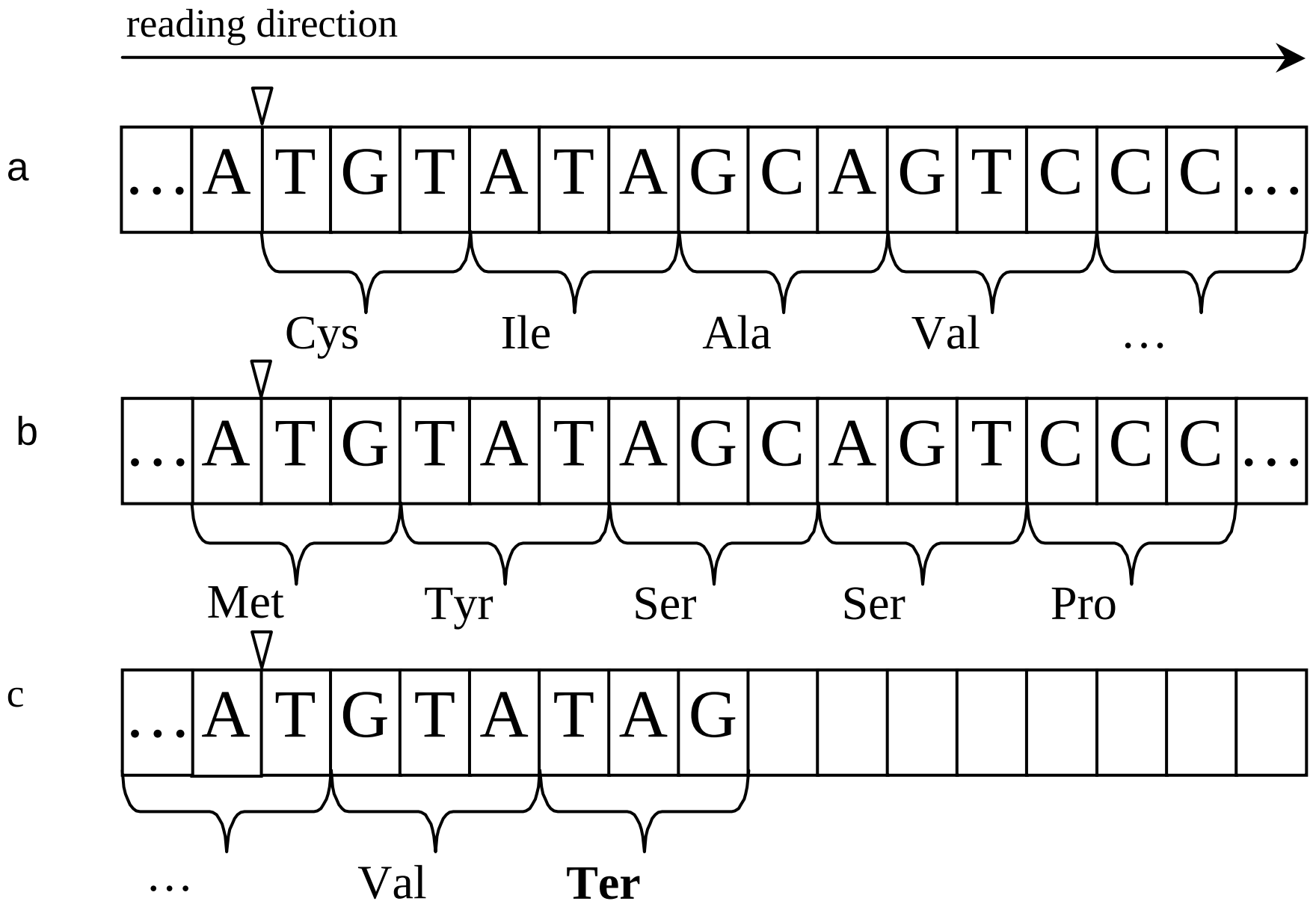

Figure 1 | Gene reading with shift.

(a) If the gene is read without a shift, it codes protein P1. However, if the reading frame is shifted by one nucleotide to the left (shift 1) (b) or to the right (shift 2) (c), then a different protein will be synthesized. But the reading with shift may result in the appearance of a stop codon as in case (c).

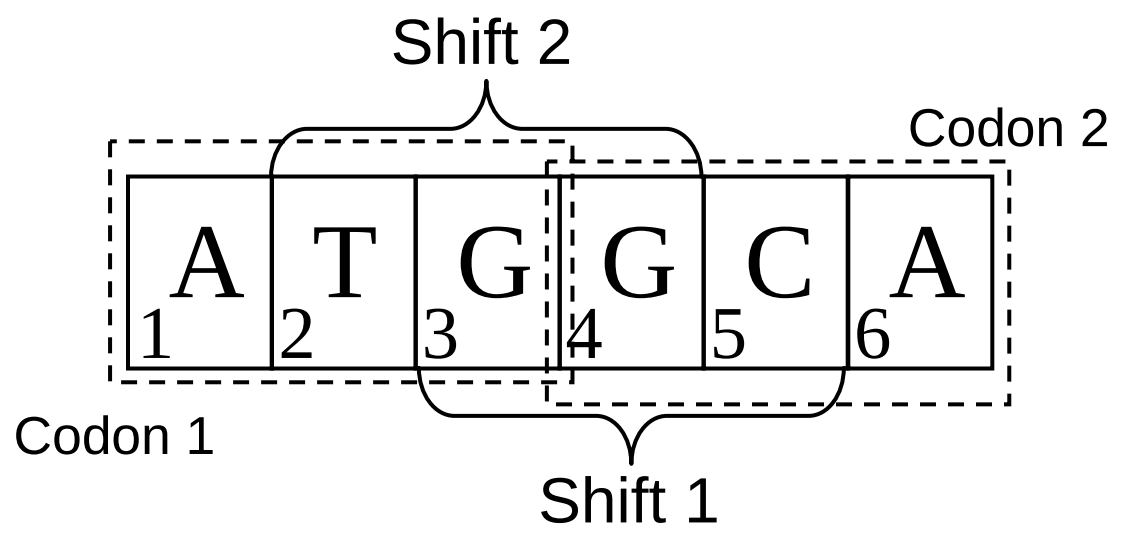

Figure 2 | A pair of codons with numbered nucleotide positions

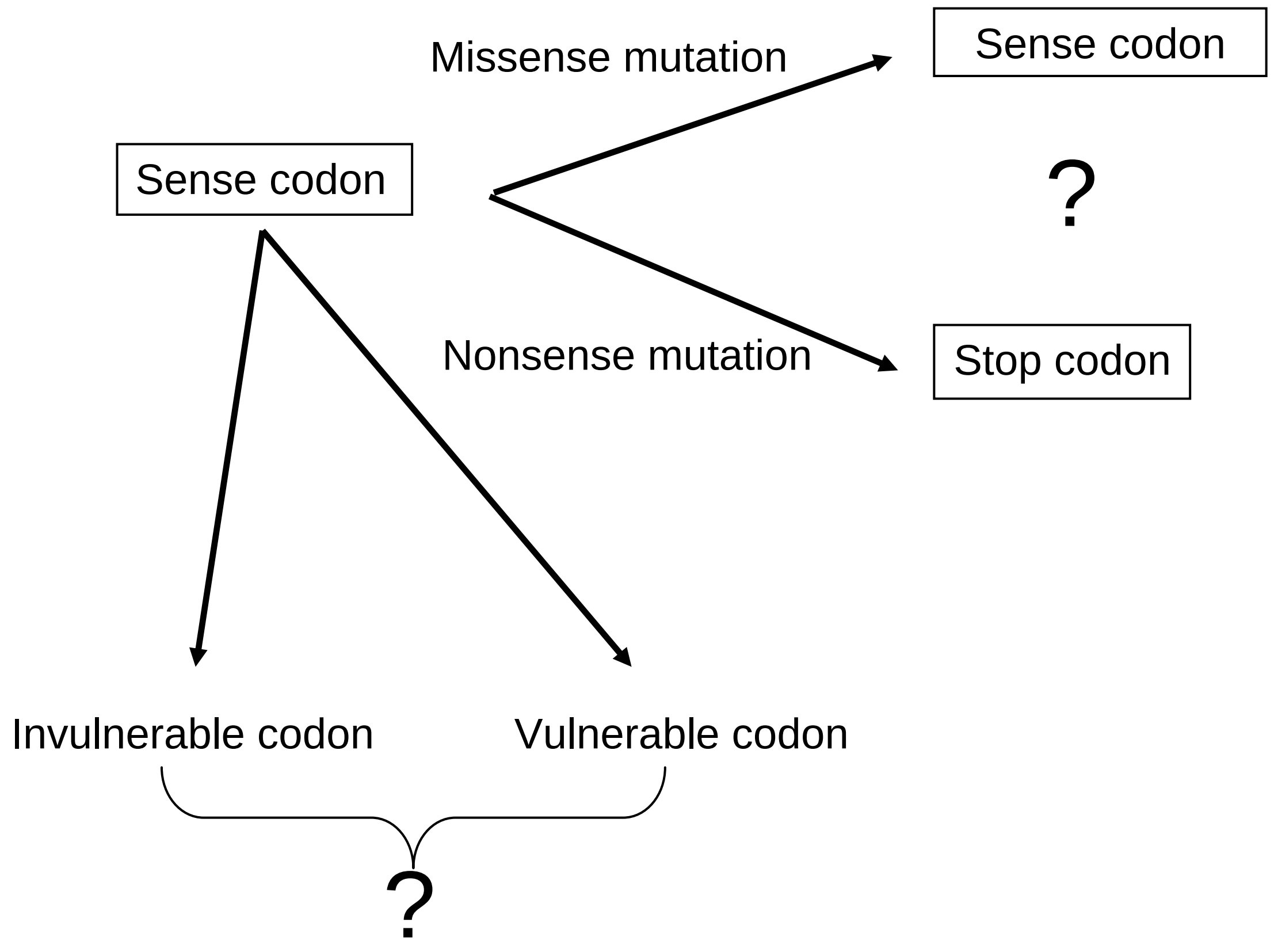

Figure 3 | Point mutations

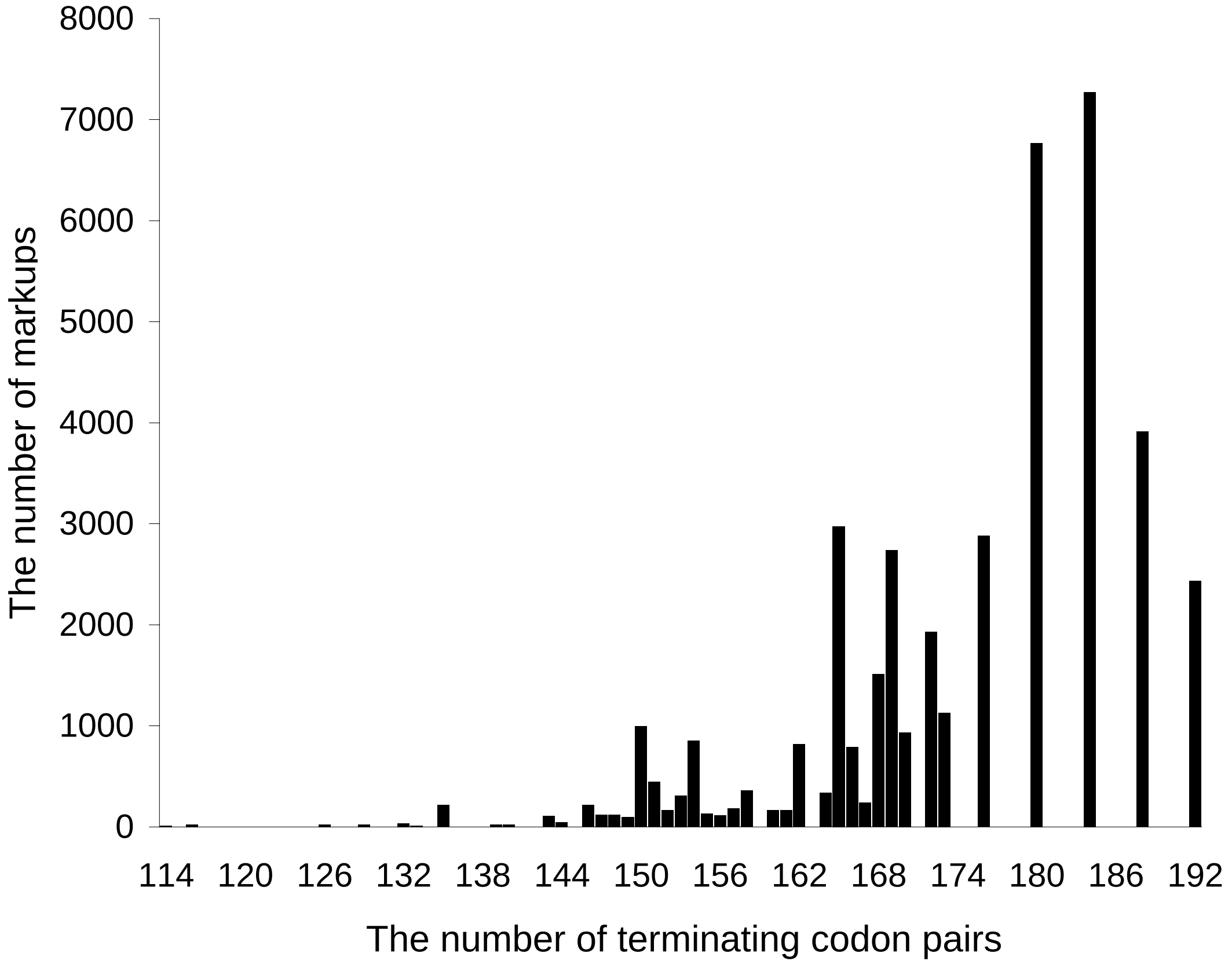

Figure 4 | The distribution of genetic code markups vs the number of terminating pairs of codons for the genetic code markups with 3 stop codons

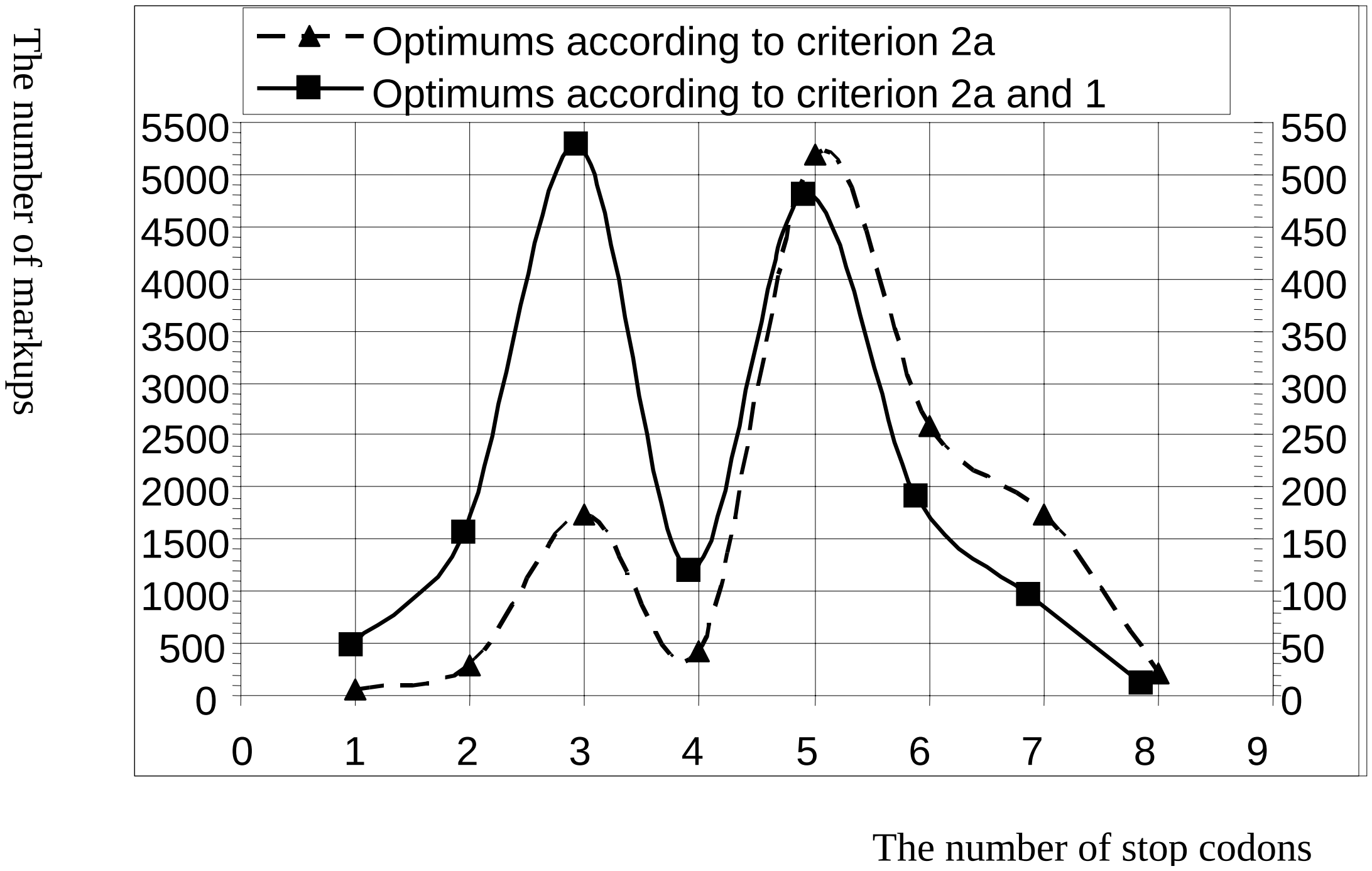

Figure 5 | Markups with the different numbers of stop codons. Powers of optimum sets.

Table 1 | The classification of genetic code markups with 3 stop codons vs the number of vulnerable codons and the number of nonsense mutations

| Variant | Group | Number of vulnerable codons | Number of nonsense mutations | Number of code markups | |
|---|---|---|---|---|---|
| | | | | Total | Rate |
| 1 | A | 18 | 23 | 1 728 | 4,15% |
| 2 | B | 19 | 21 | 192 | 0,46% |
| 3 | C | 20 | 25 | 3 456 | 8,29% |
| 4 | C | 21 | 25 | 5 184 | 12,44% |
| 5 | C | 23 | 25 | 5 184 | 12,44% |
| 6 | D | 21 | 27 | 1 152 | 2,76% |
| 7 | D | 22 | 27 | 1 728 | 4,15% |
| 8 | D | 23 | 27 | 10 368 | 24,88% |
| 9 | D | 25 | 27 | 10 368 | 24,88% |
| 10 | D | 27 | 27 | 2 304 | 5,53% |

Table 2 | Extreme properties of genetic code markups with 2, 3 and 4 stop codons

| | Number of stop codons in a genetic code markup | | |
|---|---|---|---|
| | 2 | 3 | 4 |
| Total markups | 2016 | 41664 | 635376 |
| Maximal number of terminating pairs for both shifts | 256 | 384 | 512 |
| Number of markups with maximum of terminating pairs | 564 | 2432 | 4968 |
| Minimal number of vulnerable codons | 14 | 18 | 20 |
| Number of markups with minimum of vulnerable codons | 288 | 1728 | 432 |
| Number of markups possessing both extremes | 156 | 528 | 120 |